\documentclass[aps,pra,twocolumn,amsmath,amssymb,showpacs,nofootinbib,superscriptaddress]{revtex4}

\newcommand{\bra}[1]{\langle#1|}
\newcommand{\ket}[1]{|#1\rangle}

\usepackage[dvips]{graphicx}
\usepackage{mathrsfs}

\begin{document}

\bibliographystyle{apsrev}

\title{Improving the fidelity of single photon preparation from conditional down-conversion via asymmetric multiport detection}

\author{Peter P. Rohde}
\email[]{rohde@physics.uq.edu.au}
\homepage{http://www.physics.uq.edu.au/people/rohde/}
\affiliation{Centre for Quantum Computer Technology, Department of Physics\\ University of Queensland, Brisbane, QLD 4072, Australia}

\date{\today}

\frenchspacing

% ABSTRACT
\begin{abstract}
We discuss the conditional preparation of single photons via parametric down-conversion. This technique is commonly used as a single photon source in modern quantum optics experiments. A significant problem facing this technique is the inability of present day photo-detectors to resolve photon number. This results in mixing with higher photon number terms. To overcome this several techniques have been proposed, including multi-port detection and time-division multiplexing. These techniques help approximate number resolving detection even when using non-number resolving detectors. In this paper we focus on 2-port detection, the simplest such scheme. We show that by making the 2-port device asymmetric the fidelity of prepared photons can be improved.
\end{abstract}

\pacs{}

\maketitle

\textbf{Introduction --}
Conditional preparation of single photons from parametric down-conversion has become the most popular approach to single photon generation for use in modern quantum optics and quantum information processing applications. Here a down-converter produces entangled photon pairs. One output of the down-converter is measured. When a photon is detected, with high probability there will be a photon in the other output. There is a problem with this. Namely, the down-converter produces not just single pairs, but also higher photon number terms. Most current photo-detectors are not number resolving, so these higher order terms cannot be filtered out. This places an upper bound on the fidelity of prepared states. To remedy this, several schemes have been proposed for approximating number resolving photo-detection using non-number resolving ones. These include multiport networks \cite{bib:Kok01, bib:Paul96, bib:Bartlett02, bib:Rohde05}, time division multiplexing (TDM) \cite{bib:Achilles03, bib:Achilles04, bib:Banaszek03, bib:Fitch03} and visible light photon counting modules (VLPC's) \cite{bib:Kim99, bib:Takeuchi99, bib:Bartlett02}. These schemes work by distributing the incident light field across multiple modes, which are independently measured. The probability of multiple photons arriving in the same mode drops with the number of modes, so the confidence in the measurement result increases.

In this paper we consider the simplest protocol, 2-port detection (i.e. a beamsplitter), and show that by making the beamsplitter ratio asymmetric, the fidelity of the prepared state can be improved. This comes at the expense of non-determinism. This idea was first proposed in Ref. \cite{bib:Rohde05}.

\textbf{The POVM description of measurement processes --} A measurement of photon number using a particular architecture can be characterized by a set of POVM elements. We define one POVM element for each measurement signature. A signature could be the number of photons that are measured, or, more generally, the order or timing of the detection events. For a particular signature $s$, the POVM element will take the form
\begin{equation}
\hat\Pi_s = \sum_n P_s(n)\ket{n}\bra{n},
\end{equation}
where $P_s(n)$ is the probability of detecting the signature $s$ given that $n$ photons were present. Consider the case where we detect a single photon signature. In the ideal case $P_s(n)=\delta_{n,1}$ and the projector reduces to $\ket{1}\bra{1}$. In the non-ideal case, effects like loss and dark-counts will result in non-zero values for $P_s(n\neq 1)$, and there will be undesired terms in the resultant state.

\textbf{Description of the asymmetric measurement process --}
Now let us turn our attention to the measurement scheme we are considering, a 2-port detector. Here the incident light field is distributed across two output ports using a beamsplitter. Ordinarily the splitting ratio is 50/50, but we consider the more general case where this ratio can be arbitrary, which we label $\eta_\mathrm{ref}$.

We are interested in the signature where one photon is detected. There are two distinct such signatures, so we will focus on one, where the photo-detector at the first beamsplitter output registers a click, and the other does not. We will assume detector loss occurs at rate $\eta_\mathrm{loss}$ and dark-counts at rate $\eta_\mathrm{DC}$. The probability of registering the desired signature is given by
\begin{eqnarray}
P_s(n) &=& \sum_{i=0}^n \binom{n}{i} {\eta_\mathrm{ref}}^i (1-\eta_\mathrm{ref})^{n-i} {\eta_\mathrm{loss}}^{n-i}(1-\eta_\mathrm{DC}) \nonumber\\
&\times& \left[{\eta_\mathrm{loss}}^i \eta_\mathrm{DC} + (1-{\eta_\mathrm{loss}}^i) \right].
\end{eqnarray}
There are three components in this equation. First, we sum over all possible ways in which the incident $n$ photons can be routed to the outputs. Here $i$ represents the number of reflected photons. Then, for each configuration, we multiply by the probability that the photons will give rise to the desired signature. Here we have assumed both photo-detectors are `bucket' detectors, which are only able to distinguish between no photons and some photons.

\textbf{Conditional state preparation --}
Now let us consider the application of this detection model to the conditional preparation of single photons via parametric down-conversion. A down-converter produces a two-mode entangled state of the form
\begin{equation}
\ket\psi = \sqrt{1-\chi^2} \sum_n \chi^n \ket{n}\ket{n},
\end{equation}
where $\chi$ is the down-conversion strength. Applying the measurement operator to the first mode and tracing out the measured mode we obtain
\begin{equation}
\hat\rho = (1-\chi^2) \sum_n \chi^{2n} P_s(n)\ket{n}\bra{n}.
\end{equation}
We desire to perfectly prepare the $\ket{1}$ state, so finally we calculate the fidelity, given by the magnitude of the desired term in the mixture,
\begin{equation}
F = \frac{\chi^2 P_s(1)}{\sum_n \chi^{2n} P_s(n)}.
\end{equation}

\begin{figure}[!htb]
\includegraphics[width=0.65\columnwidth]{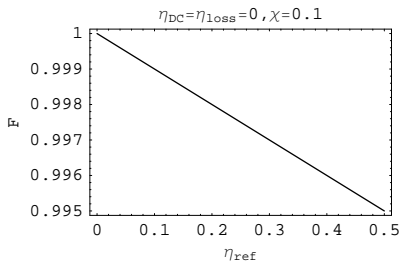}
\caption{Fidelity against splitting ratio.} \label{fig:F_vs_ref}
\end{figure}

\textbf{Results --}
A conventional 2-port device uses a 50/50 splitting ratio. In Fig. \ref{fig:F_vs_ref} we consider the ideal case where $\eta_\mathrm{DC}=\eta_\mathrm{loss}=0$ and plot the fidelity of the prepared state against the splitting ratio. Notice that as the splitting ratio is reduced the fidelity increases and approaches unity. The intuition behind this result is as follows. With two output ports and an equal splitting ratio there is a 50\% probability that when a two photon term arrives it will be confused for a single photon because they both arrive at the same output. When the splitting ratio is reduced the probability of a photon reaching the first detector also reduces. Thus, a two photon term has a low probability of causing confusion due to both photons arriving at the same output. However, there is a caveat associated with this. Namely, the improved fidelity comes at the expense of success probability. This is because we are conditioning on detecting a photon at the first beamsplitter output. The probability of this goes roughly proportional to $\eta_\mathrm{ref}$.

\begin{figure}[!htb]
\includegraphics[width=\columnwidth]{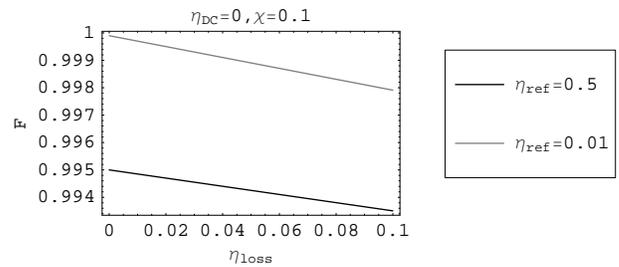}
\caption{Fidelity against loss for two different splitting ratios.} \label{fig:F_vs_loss}
\end{figure}

In Fig. \ref{fig:F_vs_loss} we plot the fidelity against $\eta_\mathrm{loss}$ for two different values of $\eta_\mathrm{ref}$. This demonstrates that using an asymmetric beamsplitter ratio is beneficial also in the presence of loss.

\begin{figure}[!htb]
\includegraphics[width=0.7\columnwidth]{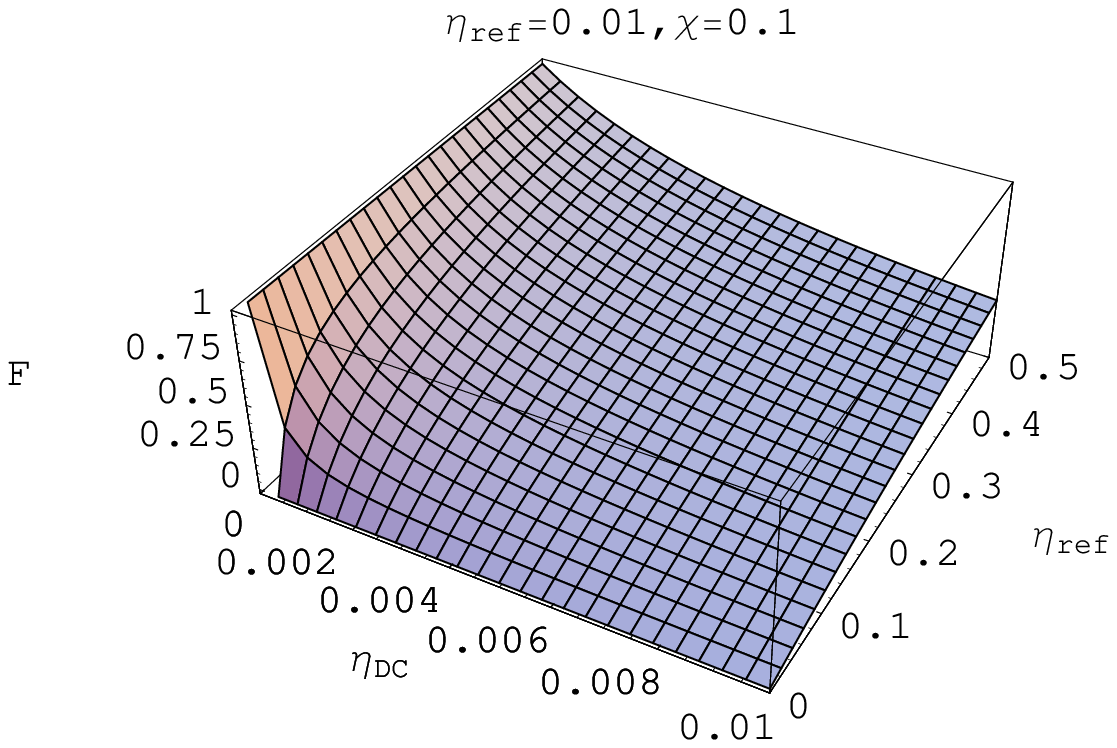}\\
\includegraphics[width=0.7\columnwidth]{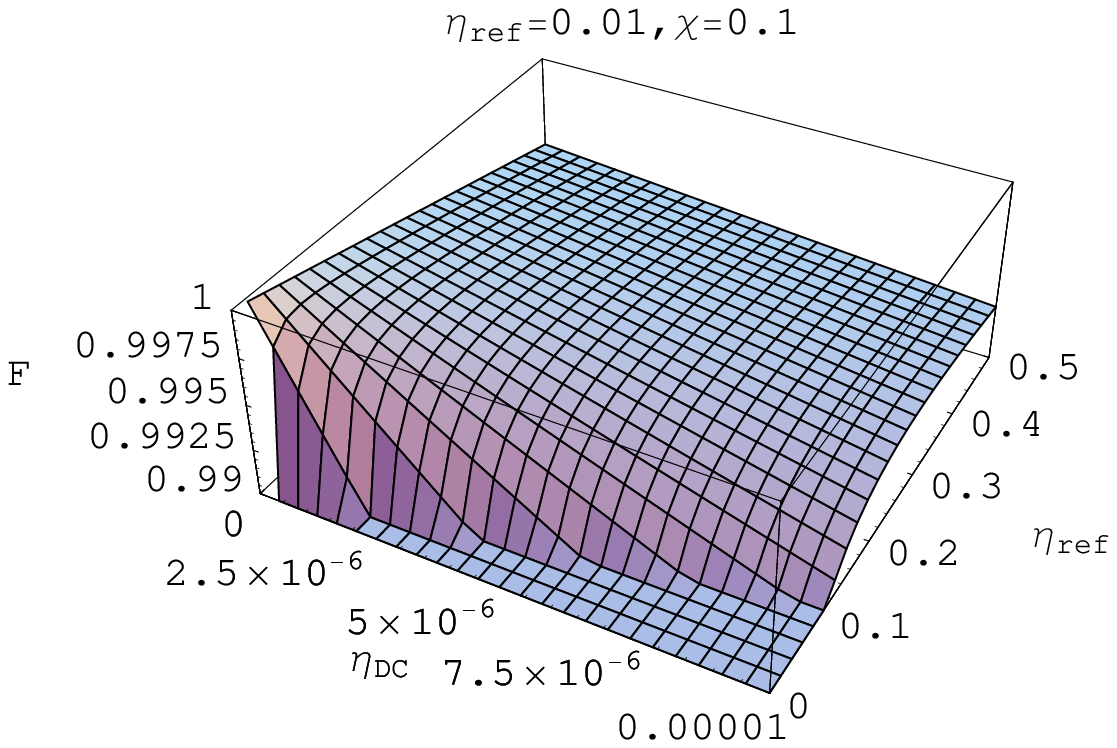}
\caption{Fidelity against dark-count rate and splitting ratio. $\eta_\mathrm{loss}=0$.} \label{fig:F_vs_DC_ref}
\end{figure}

In Fig. \ref{fig:F_vs_DC_ref} we plot fidelity against both $\eta_\mathrm{DC}$ and $\eta_\mathrm{ref}$. Notice that the fidelity drops extremely rapidly with dark-counts. Also, in the first inset, notice that it appears to be beneficial to increase the coupling rate, which apparently contradicts our earlier findings. In the second inset we zoom in on the region where dark-count rates are extremely low. Here we in fact see that there is an optimal value of $\eta_\mathrm{ref}$, depending on the dark-count rate. Notice that on the axis where $\eta_\mathrm{DC}=0$ the plot reduces to the linear plots we observed earlier.

\textbf{Summary -- }
We have discussed a variation on the well known 2-port approach to photon number measurement. Specifically we considered the application of this scheme to conditional state preparation via parametric down-conversion. Our analysis includes the effects of loss and dark-counts. Our results indicate that using an asymmetric beamsplitter can benefit this scheme. However, this comes at the expense of success probability. Also, it is important to realize that the benefits begin to be realized in the regime of 99.5\% fidelity, which is beyond what is presently achievable. Thus, our results are unlikely to be of present practical applicability given the large number of other factors that currently limit fidelity.

% ACKNOWLEDGMENTS
\begin{acknowledgments}
We thank James Webb, Elanor Huntington and Timothy Ralph for helpful discussions. This work was supported by the Australian Research Council and Queensland State Government. We acknowledge partial support by the DTO-funded U.S. Army Research Office Contract No. W911NF-05-0397.
\end{acknowledgments}

% BIBLIOGRAPHY
\bibliography{paper}

\end{document}